\documentclass[aps,prd,twocolumn,superscriptaddress,preprintnumbers,floatfix,nofootinbib,notitlepage,showkeys]{revtex4-1}

\usepackage[utf8]{inputenc}
\usepackage{cancel}
\usepackage{float}
\usepackage{graphicx}
\usepackage{hyperref}
\usepackage{latexsym}
\usepackage{amsmath}
\usepackage{amssymb}
\usepackage{bbm}

\usepackage{ulem}
\usepackage{pdfsync}
\usepackage{epsfig}
\usepackage{epstopdf}
\usepackage{subfigure}
\usepackage{color}
\usepackage{comment}
\usepackage{slashed}

\def\beq{\begin{equation}}
\def\eeq{\end{equation}}
\def\baq{\begin{eqnarray}}
\def\eaq{\end{eqnarray}}

\newcommand{\be}{\begin{equation}} 
\newcommand{\ee}{\end{equation}}
\newcommand{\bea}{\begin{equation}\begin{aligned}} 
\newcommand{\eea}{\end{aligned}\end{equation}}

\newcommand{\bmp}{\noindent\begin{minipage}{16cm}}
\newcommand{\emp}{\end{minipage}\vskip 7mm} 
\def\lsim{\mathrel{\raise.3ex\hbox{$<$\kern-.75em\lower1ex\hbox{$\sim$}}}}
\def\gsim{\mathrel{\raise.3ex\hbox{$>$\kern-.75em\lower1ex\hbox{$\sim$}}}}

\newcommand{\intron}[1]{}

\begin{document}
\title{Standard model Higgs field and hidden sector cosmology}

\author{Tommi Tenkanen}
\email{ttenkan1@jhu.edu}
\affiliation{Department of Physics and Astronomy, Johns Hopkins University, \\
Baltimore, MD 21218, USA}

\begin{abstract}
We consider scenarios where the inflaton field decays dominantly to a hidden dark matter (DM) sector. By studying the typical behavior of the Standard Model (SM) Higgs field during inflation, we derive a relation between the primordial tensor-to-scalar ratio $r$ and amplitude of the residual DM isocurvature perturbations $\beta$ which is typically generated if the DM is thermally decoupled from the SM sector. We consider different expansion histories and find that if the Universe was radiation- or matter-dominated after inflation, a future discovery of primordial DM isocurvature will rule out all simple scenarios of this type because generating observable $\beta$ from the Higgs is not possible without violating the bounds on $r$. Seen another way, the Higgs field is generically not a threat to models where both the inflaton and DM reside in a decoupled sector. However, this is not necessarily the case for an early kination-dominated epoch, as then the Higgs can source sizeable $\beta$. We also discuss why the Higgs cannot source the observed curvature perturbation at large scales in any of the above cases but how the field can still be the dominant source of curvature perturbations at small scales.
\end{abstract}

%
\maketitle

%

\section{Introduction}
\label{intro}

The existence of the Standard Model (SM) Higgs boson is by now a well-established fact. However, currently only little is known about the dynamics of the Higgs field in the early Universe and how the physics at very high energies are affected by the properties of the Higgs. Examples of scenarios where the field has been suggested to play an important role include those where during cosmic inflation, an early epoch of exponential expansion in the early Universe, the Higgs field either assumed the role of the inflaton, i.e. was responsible for causing the exponential expansion, \cite{Bezrukov:2007ep,Bauer:2008zj,Kobayashi:2011nu,Rasanen:2018ihz} or remained as an energetically subdominant {\it spectator} field which did not take part in driving inflation and left no imprints on inflationary observables \cite{Enqvist:2013kaa,Hook:2014uia,Enqvist:2014bua,Espinosa:2015qea}. However, even then the field may have had an important effect on the physics after inflation. For example, the field may have affected post-inflationary reheating \cite{Freese:2017ace}, initial conditions for baryo- or leptogenesis \cite{Enqvist:2014zqa,Kusenko:2014lra,Yang:2015ida,Pearce:2015nga,Wu:2019ohx}, dark matter production and electroweak symmetry breaking \cite{Enqvist:2014zqa,Cosme:2018wfh}, or generation of curvature perturbation \cite{DeSimone:2012qr,Kunimitsu:2012xx,Figueroa:2016dsc}.

Assuming the SM is a valid theory of the SM particle properties and interactions up to high energies, the known properties of the SM Higgs provide for a powerful probe to the physics of the very early Universe or, equivalently, the physics at very high energies. To demonstrate that, in this paper we consider scenarios where inflation was driven by a field other than the Higgs, an inflaton, and where, after inflation, the inflaton decayed dominantly to a hidden sector which was thermally decoupled from the visible SM sector. By utilizing the known properties of the Higgs field, we place constraints on this type of scenarios which apply regardless of how weakly the hidden sector couples to the SM. The scenario is well-motivated, as it has recently been suggested that in order for the electroweak vacuum to remain stable and/or the quantum corrections due to the Higgs ensure flatness of the inflaton potential at high energies, the coupling between the inflaton and the Higgs indeed has to be very small \cite{Gross:2015bea,Ema:2016kpf,Kohri:2016wof,Enqvist:2016mqj,Ema:2017ckf,Rusak:2018kel}, in some cases even smaller than $g\sim 10^{-8}$ for an operator of the type $g\phi^2h^2$, where $h$ is the Higgs and $\phi$ the inflaton \cite{Rusak:2018kel}. If that is the case, then it is likely that the SM and the hidden sector did not enter into thermal equilibrium with each other in the early Universe \cite{Petraki:2007gq,Merle:2013wta,Enqvist:2014zqa,Kahlhoefer:2018xxo}.

Similar to Refs. \cite{Berlin:2016vnh,Tenkanen:2016jic,Berlin:2016gtr,Heurtier:2019eou}, we assume that the particle(s) that constitute the observed dark matter (DM) component in the Universe also reside in the hidden sector together with the other decay products of the inflaton\footnote{See Refs. \cite{Bastero-Gil:2015lga,Tenkanen:2016twd,Almeida:2018oid,Choi:2019osi} for similar scenarios where the DM {\it is} the inflaton.}. We assume that some component in the hidden sector, e.g. the lightest of the hidden sector particles, came to dominate the energy density at some epoch in the early Universe and decayed into the SM sector and thus reheated it only after the comoving DM density was generated. As shown in the above works, due to the effective dilution of the DM abundance at reheating, the DM component can in such scenarios be as heavy as $10^{10}$ GeV while still being a thermal relic. This is in contrast to the standard upper limit $\sim 100$ TeV imposed by unitarity in a scenario where the DM component is a thermal relic from the SM sector heat bath \cite{Griest:1989wd}.

One may think that the only requirement for such a scenario involving hidden sector DM and particle decays is that the SM degrees of freedom must come to dominate the energy density of the Universe before Big Bang Nucleosynthesis at $T\simeq \mathcal{O}(1)$ MeV \cite{Kawasaki:2000en,Hannestad:2004px,Ichikawa:2005vw,DeBernardis:2008zz}. However, as we will show in this work, also the observational constraints on DM {\it isocurvature} perturbations can play an important role in determining viability of different hidden sector models. By DM isocurvature one refers to the possibility that perturbations in the local DM energy density do not coincide with those in the SM radiation. Because in scenarios involving decoupled hidden sectors typically {\it both} the inflaton and the SM Higgs acquire fluctuations during inflation, there are no one but two sources of SM radiation energy density. If the comoving DM density froze before the lightest hidden sector state decayed into the SM sector, then the local perturbations in DM energy density indeed will not generically coincide with those in SM radiation, as DM was not sourced by the Higgs but only the inflaton. 

The current observational limits on primordial DM isocurvature perturbations obtained from measurements of the Cosmic Microwave Background radiation (CMB) are very stringent, i.e. the observations are consistent with the absence of primordial isocurvature perturbations \cite{Aghanim:2018eyx,Akrami:2018odb}. The question therefore is: do the fluctuations the Higgs typically acquires during inflation pose a threat to scenarios where DM resides in a hidden sector -- or can they even be a virtue? The latter possibility is particularly interesting, as the analysis of the CMB by the Planck collaboration finds more smoothing of the so-called acoustic peaks of the CMB than predicted by the usual $\Lambda$CDM model. The collaboration has suggested that this may be explained by DM isocurvature perturbations with a blue-tilted power spectrum \cite{Aghanim:2018eyx,Akrami:2018odb} (see, however, Ref. \cite{Domenech:2019cyh} for an alternative explanation). There are plenty of well-motivated scenarios where a cold DM component generates a primordial isocurvature mode, such as axion DM \cite{Marsh:2015xka,Graham:2018jyp,Guth:2018hsa,Schmitz:2018nhb} or a spectator DM consisting of a scalar \cite{Peebles:1999fz,Nurmi:2015ema,Kainulainen:2016vzv,Bertolami:2016ywc,Heikinheimo:2016yds,Cosme:2017cxk,Enqvist:2017kzh,Cosme:2018nly,Alonso-Alvarez:2018tus,Markkanen:2018gcw,Tenkanen:2019aij,AlonsoAlvarez:2019cgw}, a vector \cite{Graham:2015rva,AlonsoAlvarez:2019cgw}, or a fermion component \cite{Kainulainen:2016vzv,Heikinheimo:2016yds}. In many cases, the isocurvature power-spectrum is naturally blue-tilted \cite{Peebles:1999fz,Graham:2015rva,Bertolami:2016ywc,Cosme:2017cxk,Cosme:2018nly,Alonso-Alvarez:2018tus,Markkanen:2018gcw,Tenkanen:2019aij} and therefore can potentially mimic the smoothing effect of lensing and/or lead to enhanced structure formation \cite{Graham:2015rva,Alonso-Alvarez:2018tus,Tenkanen:2019aij}.

However, in this paper we concentrate on an alternative way to generate the DM isocurvature perturbations, the one where the source is the SM Higgs. We show that in scenarios where the DM resides in a decoupled hidden sector isocurvature perturbations are generically generated and there is a simple relation between the energy scale of cosmic inflation and amplitude of the DM isocurvature perturbations at large scales. The present work generalizes the results originally presented in Ref. \cite{Tenkanen:2016jic}, as here we consider different expansion histories and compute the primordial DM isocurvature spectrum more carefully. 

We find, in particular, that in the standard case where the Universe was radiation-dominated after inflation or when it underwent an early matter-dominated epoch, the Higgs field does not typically pose a threat to models where both inflaton and DM reside in a decoupled hidden sector, and any evidence for primordial isocurvature will unambiguously point to the existence of a non-thermal DM component. However, as we demonstrate, this is not necessarily the case for other post-inflationary expansion histories. We also show why the fluctuations in the Higgs field cannot source the observed curvature perturbation at large scales in any of the cases we consider but how the field can still be the dominant source of curvature perturbations at small scales.

The paper is organized as follows: in Sec. \ref{dynamics} we discuss the dynamics of the Higgs and the hidden sector both during and after inflation, in Sec. \ref{isocSection} we discuss perturbations in energy densities of different components and derive the relation between the tensor-to-scalar ratio and amplitude of primordial DM isocurvature perturbations, and in Sec. \ref{results} discuss the results and observational consequences on hidden sector models in different cases. In Sec. \ref{conclusions} we conclude.

%

\section{Dynamics of the Higgs and the hidden sector}
\label{dynamics}

In this Section, we discuss the dynamics of the Higgs and a decoupled hidden sector both during and after inflation. We begin by discussing the hidden sector dynamics in Sec. \ref{sec:hidden_sector}, and then discuss the Higgs dynamics during and after inflation in Sec. \ref{sec:higgs_inf} and Sec. \ref{sec:higgs_postinf}, respectively.

\subsection{Dynamics of the hidden sector}
\label{sec:hidden_sector}

As discussed in Sec. \ref{intro}, we assume that the field responsible for inflation, the inflaton, resides in a hidden sector together with the field(s) that constitute the observed DM abundance, as well as with the field that eventually decays into the SM sector. We assume that after inflation the inflaton decays solely to the hidden sector and imprints its primordial fluctuation spectrum into perturbations in the energy density of the hidden sector plasma in the usual way. We assume the hidden sector then remains as the energetically dominant sector up until to a point when the lightest state decays into the SM sector, thus making it the dominant energy component. We assume this happens only after the comoving DM number density has frozen to a constant value but before BBN. As will become evident in Sec. \ref{results}, these assumptions are motivated by the fact that they give in most cases the maximum effect the Higgs field can have on observables. This is exactly what we want to concentrate on when investigating if an energetically subdominant Higgs can be a threat to the success of the scenario under any circumstances.

We approximate the Friedmann equation by
\begin{equation}
3H^2 M_{\rm P}^2 \simeq 
\begin{cases}
\rho_{\rm hid} & \quad a<a_{\rm reh} \\
\rho_{\rm SM} & \quad a\geq a_{\rm reh}
\end{cases}
\end{equation}
where $H(a)$ is the Hubble parameter, $M_{\rm P}$ is the reduced Planck mass, and $\rho_{\rm hid}(a)$ and $\rho_{\rm SM}(a)$ are the hidden and SM sector energy densities, respectively. We work in the sudden decay approximation and denote the time when the hidden sector finally decays into the SM by $a_{\rm reh}$. Evolution of the scale factor $a$ is governed by the hidden sector until the SM radiation becomes the dominant energy density component and the usual Hot Big Bang era begins. This is what we call 'reheating'. 

Before reheating, evolution of $a$ can differ from the one induced by the usual radiation-domination. We assume that before reheating the hidden sector fluid governing the total energy density had a barotropic equation of state controlled by an effective equation of state parameter $w=\rho/p$ from the end of inflation to the time of reheating, where $p$ is the pressure of the hidden sector fluid, so that
\be
H = H_*\left(\frac{a_*}{a}\right)^{\frac{3(1+w)}{2}} ,
\ee
where $H_*$ and $a_*$ are the values of the Hubble scale and scale factor at the end of inflation, respectively. In the following, we normalize the scale factor such that at the beginning of post-inflationary era $a_*=1$, whereas in single-field models of inflation, the scale of inflation $H_*$ can be expressed with the observable tensor-to-scalar ratio $r\equiv \mathcal{P}_t/\mathcal{P}_\zeta$ as
\be
\label{Hr}
\frac{H_*}{M_{\rm P}} = \sqrt{\frac{\pi^2\mathcal{P}_{\zeta}r}{2}},
\ee
where $r$ is bound from above by the CMB measurements, $r<0.06$ (at the $2\sigma$ level) \cite{Ade:2018gkx} and the power spectrum of the curvature perturbation $\zeta$ has the amplitude $\mathcal{P}_{\zeta}\simeq 2.1\times 10^{-9}$ (at the $1\sigma$ level) \cite{Akrami:2018odb}. The curvature perturbation will be defined in Eq. \eqref{zeta}. Here we also used $\mathcal{P}_t=8(H_*/2\pi)^2/M_{\rm P}^2$ for the amplitude of tensor perturbations (see e.g. Ref. \cite{Riotto:2002yw}). The result \eqref{Hr} will become useful later.

\subsection{Dynamics of the Higgs field during inflation}
\label{sec:higgs_inf}

It is well-known that scalar fields $\sigma$ which are light, $d^2V(\sigma)/d\sigma^2 < H^2_*$, and energetically subdominant, $\rho_\sigma \ll 3H_*^2M_{\rm P}^2$, typically acquire large fluctuations during cosmic inflation. The magnitude of such fluctuations in one Hubble time is proportional to the inflationary scale $H_*$ \cite{Starobinsky:1986fx,Starobinsky:1994bd}. In particular, this is the case for the SM Higgs field \cite{Enqvist:2013kaa}, unless the electroweak vacuum is metastable, the Higgs field is coupled to operators which make it heavy during inflation, or if it is coupled strongly and non-minimally to gravity, rendering our treatment inadequate or preventing the field from gaining such large fluctuations \cite{EliasMiro:2011aa,Kobakhidze:2013tn,Fairbairn:2014zia,Enqvist:2014bua,Herranen:2014cua,Hook:2014uia,Herranen:2015ima,Espinosa:2015qea}. In this paper we do not consider these possibilities but assume that the Higgs was indeed light enough during inflation to acquire fluctuations and that its potential remains stable up to the highest scales considered in this work.

Assuming the Higgs potential during inflation
\be
V(h) = \frac{\lambda}{4}h^4,
\ee
the one-point probability distribution of the Higgs field at the end of inflation is given by~\cite{Starobinsky:1986fx,Starobinsky:1994bd}
\be
\label{Peq}
P(h) =  N\exp\left(-\frac{8\pi^2V(h)}{3H_*^4}\right),
\ee
where $N$ is a normalization factor ensuring total probability equal to unity. The distribution \eqref{Peq} describes an effective position-dependent Higgs condensate with a typical field value \cite{Enqvist:2013kaa}
\be 
\label{hstar}
h_{*}\equiv \sqrt{\langle h^2\rangle} = \left(\frac{3}{2\pi^2}\right)^{1/4}\sqrt{\frac{\Gamma(\frac{3}{4})}{\Gamma(\frac{1}{4})}} \frac{H_*}{\lambda^{1/4}} \equiv A\lambda^{-1/4}H_*,
\ee 
in a patch the size of the horizon at the end of inflation. Here the prefactor $A\simeq 0.363$ is defined for later purposes. Thus, the average energy density of the Higgs field at the end of inflation is
\be
\rho_h=\langle V(h)\rangle=\frac{3H_*^4}{32\pi^2},
\ee
which shows that the Higgs is indeed subdominant to the total energy density by a factor $H_*^2/M_{\rm P}^2$. There is slight ambiguity in the result \eqref{hstar} related to the inflationary background, as $H_*$ is not strictly constant during inflation and relaxation of the Higgs' probability distribution to its equilibrium state takes a finite time \cite{Enqvist:2012xn,Hardwick:2017fjo}. Here we assume that $H_*$ was constant during inflation and the equilibrium state was reached prior to the end of inflation, so that Eq. \eqref{hstar} indeed gives the typical initial condition for post-inflationary dynamics. This is a well-motivated assumption, as in slow-roll inflation $|\dot{H}|=\epsilon H^2$ with $\epsilon \ll 1$ the Hubble scale changes only very little during inflation and, on the other hand, the equilibrium state characterized by Eq. \eqref{Peq} is reached in $N\sim 1/\sqrt{\lambda}$ e-folds regardless of the initial value of the field \cite{Enqvist:2012xn}.

\subsection{Dynamics of the Higgs field after inflation}
\label{sec:higgs_postinf}

The post-inflationary evolution of the Higgs condensate has been studied in detail in Refs. \cite{Enqvist:2013kaa,Enqvist:2014zqa,Enqvist:2014tta,Figueroa:2015rqa,Enqvist:2015sua,Lozanov:2016pac}. The Higgs field remains initially frozen at the value $h$ it had at the end of inflation in a given patch, and begins to oscillate about its minimum after it becomes massive at $H=H_{\rm osc}$, where
\be
\label{Hosc}
H_{\rm osc}^2(h)=3\lambda h^2.
\ee
Typically, the field decays into SM particles in $\mathcal{O}(1)$ e-folds. The produced particles form a heat bath whose energy density $\rho^h_\gamma$ scales down as radiation
\be
\label{equ:gammarho}
\rho^h_\gamma=\frac{1}{4}\lambda h^4\left(\frac{a_{\rm osc}(h)}{a}\right)^4,
\ee
where $a_{\rm osc}(h)$ is the scale factor at the time when the Higgs started to oscillate. In the following, we assume that the Higgs' oscillations always begin before reheating.

Because the Higgs only constitutes a small fraction of the total energy density, in our scenario the SM particles are initially subdominant to the hidden sector heat bath which was generated by decay of the energetically dominant inflaton field, as discussed in Sec. \ref{sec:hidden_sector}. This means that at a time when the Hubble rate is $H$, the energy density of the Higgs decay products (\ref{equ:gammarho}) is
\be
\label{equ:gammascaling}
\rho^h_\gamma=\frac{1}{4}\lambda h^4\left(\frac{H^2}{H^2_{\rm osc}(h)}\right)^\frac{4}{3(1+w)}
\propto |h|^{\frac{4}{3}\frac{1+3w}{1+w}}
,
\ee
where $h$ is the local value of the Higgs field at the end of inflation and we used Eq. \eqref{Hosc} for $H_{\rm osc}$. For example, in the case of the radiation equation of state, $w=1/3$, we obtain $\rho^h_\gamma=H^2h^2/12$. Assuming that the SM particles thermalize instantaneously upon their production, the Hubble parameter at the time of reheating
$H_{\rm reh}$ can be expressed in terms of the reheating temperature
\be
\label{HandTreh}
H(a_{\rm reh}) = \sqrt{\frac{\pi^2 g_*(T_{\rm reh})}{90}}\frac{T_{\rm reh}^2}{M_{\rm P}} \equiv B\frac{T_{\rm reh}^2}{M_{\rm P}} ,
\ee
where $g_*(T_{\rm reh})$ is the effective number of degrees of freedom at the time of reheating and $B$ has again been defined for later purposes. For the SM particle content $g(T_{\rm reh})=106.75$ for $T_{\rm reh}\gg 100$ GeV, so $B\simeq 3.42$.

The above results are straightforward to generalize to scenarios where for example prior to the time when the inflaton decayed into hidden sector particles the effective equation of state parameter took a value different from the one during subsequent evolution of the Universe, $w_{\rm dec}\neq w$. In this paper we do not consider this possibility. In any case, using Eq. \eqref{Hr} we can derive an absolute, model-independent upper bound on the reheating temperature in terms of observables,
\be
\label{Trehupper}
\frac{T_{\rm reh}}{M_{\rm P}} \leq \left(\frac{\pi^2\mathcal{P}_\zeta r_{\rm max}}{2B^2}\right)^{\frac{1}{4}} \simeq 0.003 ,
\ee
corresponding to $T_{\rm reh}^{\rm max} \simeq 6.6\times 10^{15}\, {\rm GeV}$. Here we used $\mathcal{P}_\zeta=2.1\times 10^{-9}$, $r_{\rm max}=0.06$, and $B=3.42$. However, the assumption that in our scenarios the Higgs' oscillations begin before reheating gives a slightly more stringent limit for the reheating temperature
\be
\label{Trehupper2}
\frac{T_{\rm reh}}{M_{\rm P}} < \lambda^{1/8}\left(\frac{A^2\pi^2\mathcal{P}_\zeta r}{2B^2}\right)^{\frac{1}{4}} \simeq 0.004 r^{1/4}\lambda^{1/8} ,
\ee
which follows from Eqs. \eqref{Hosc} and \eqref{HandTreh} by requiring $H(a_{\rm reh}) < H(a_{\rm osc})$.

%

\section{Perturbations in energy densities}
\label{isocSection}

If all fluid components were in thermal equilibrium in the early Universe, there would be no isocurvature perturbations. Thus, in case of a discovery of such perturbations, all scenarios with perfect thermal equilibrium will be ruled out. Because primordial DM isocurvature is at the moment detectable only at very large scales, observable isocurvature perturbations must have been sourced during inflation. As single-field models cannot source isocurvature \cite{Weinberg:2004kr}, their detection would point to non-trivial multifield dynamics during inflation. There would then be two options: either the Higgs is the inflaton \cite{Bezrukov:2007ep,Bauer:2008zj,Kobayashi:2011nu,Rasanen:2018ihz} and the isocurvature perturbatios are generated by some other spectator field, for example an axion \cite{Marsh:2015xka,Graham:2018jyp,Guth:2018hsa,Schmitz:2018nhb} or some other field such as a scalar \cite{Peebles:1999fz,Nurmi:2015ema,Kainulainen:2016vzv,Bertolami:2016ywc,Heikinheimo:2016yds,Cosme:2017cxk,Enqvist:2017kzh,Cosme:2018nly,Alonso-Alvarez:2018tus,Markkanen:2018gcw,Tenkanen:2019aij, AlonsoAlvarez:2019cgw} or vector DM \cite{Graham:2015rva,AlonsoAlvarez:2019cgw}, or inflation is driven by something else than the Higgs and the Higgs is a spectator sourcing isocurvature. 

In this section we study the latter case above, where the DM isocurvature perturbations are sourced by the primordial {\it Higgs} condensate. During inflation, the Higgs acquired fluctuations, i.e. there were patches with $h$ different from Eq. \eqref{hstar}, which were {\it a priori} completely uncorrelated with perturbations in the hidden sector. We assume perturbations within the hidden sector are completely adiabatic, i.e. all particle species in the hidden sector share the same perturbations, and that the primordial fluctuations in the Higgs were imprinted into the SM radiation upon the decay of the Higgs condensate. Therefore, provided that the comoving DM number density freezes to a constant value before reheating and never thermalized with the visible SM sector, the non-zero fluctuations in the Higgs field generate a residual isocurvature perturbation\footnote{For 'adiabatic' perturbations $S=0$.} between the DM and SM radiation sector
\be
\label{isocurvature}
S\equiv \frac{3}{4}\frac{\delta\rho_{\gamma}}{\rho_{\gamma}}  - \frac{\delta\rho_{c}}{\rho_{c}} ,
\ee
where perturbations of a fluid component $i=\gamma, c$ are defined as $\delta\rho_i\equiv \rho_i(x) -\langle \rho_i\rangle$.

To see at a more detailed level how the DM isocurvature arises and to make use of the observational constraints, we divide the final energy density of the baryon-photon fluid, $\rho_\gamma$, into a part which was sourced by the Higgs condensate, $\rho_{\gamma}^{h}$, and to a part which was sourced by the hidden sector at the time of SM reheating, $\rho_{\gamma}^{H}$, so that $\rho_{\gamma} = \rho_{\gamma}^{h} + \rho_{\gamma}^{H}$. We assume that first the comoving DM density freezes and then the hidden sector decays into SM radiation, so that both components will inherit their perturbation spectra from the hidden sector, $(3/4)\delta\rho_{\gamma}^{H}/\rho_{\gamma}^{H} = \delta\rho_c/\rho_c \propto \zeta$, where 
\begin{equation}
\label{zeta}
\zeta \equiv -\Phi -H\frac{\delta\rho}{\dot{\rho}}
\end{equation}  
is the total curvature perturbation on the uniform energy density hypersurface, $\Phi$ is the gravitational potential in the longitudinal gauge, $\rho$ is the total energy density, and we assume that the continuity equation $\dot{\rho}=3H(1+w)\rho$ holds at all times, i.e. the eventual decay of the hidden sector to the SM is instantaneous. Thus, the isocurvature perturbation \eqref{isocurvature} becomes
\be
\label{S}
S = \frac{\rho_\gamma^h}{\rho_\gamma^h + \rho_\gamma^H}\left(\frac34\frac{\delta\rho_\gamma^h}{\rho_\gamma^h} - \frac{\delta\rho_c}{\rho_c}\right) .
\ee

After the hidden sector has decayed into SM radiation, the perturbations in the two sectors are separately conserved as both sectors consist of non-interacting fluids with a constant equation of state \cite{Wands:2000dp}. The isocurvature perturbations imprinted on the CMB can thus be evaluated right at the point when the hidden sector decays. Because in this scenario DM is sourced by the hidden sector and radiation is sourced by both the hidden sector and also in part by the Higgs, the resulting DM isocurvature perturbation is correlated with the curvature perturbation $\zeta$.

The power spectrum of $S$ is defined in the standard manner as a Fourier transform
\be
{\cal P}_S(k)=\frac{k^3}{2\pi^2}\int d^3x e^{i\vec{k}\cdot\vec{x}}\langle S(0)S(\vec{x})\rangle
\label{eq:P}\,.
\ee
and can in our case be written as
\be
\label{equ:PS}
\mathcal{P}_{S} = \left(\frac{\rho_{\gamma}^{h}}{\rho_{\gamma}^{h}+\rho_{\gamma}^{H}}\right)^2\left(1+ \left(\frac{3}{4}\right)^2\frac{\mathcal{P}_{\delta h}}{\mathcal{P}_{\zeta}}\right)\mathcal{P}_{\zeta}\,, 
\ee
where we used $\langle \delta\rho^{h}_{\gamma} \delta\rho_c\rangle=0$ and denoted by
$\mathcal{P}_{\delta h}$ the power spectrum of the radiation energy density perturbation sourced by the Higgs condensate,
$\delta h\equiv\delta\rho^h_\gamma/\rho^h_\gamma$. Because
\begin{equation}
\label{perturbations_higgs}
\frac{\delta\rho^h_\gamma}{\rho^h_\gamma} = \frac{\delta\left( |h|^{\frac{4}{3}\frac{1+3w}{1+w}}\right)}{|h|^{\frac{4}{3}\frac{1+3w}{1+w}}} \equiv \frac{\delta f(h)}{f(h)},
\end{equation}
as given by Eq. \eqref{equ:gammascaling}, we can compute the the two-point correlator of the energy density perturbations as a function of the $n$-point correlators of the field 
\begin{equation}
\langle \delta_h(0)\delta_h(r) \rangle =\frac{\langle f(h(0))f(h(r))\rangle-\langle f(h)\rangle^2}{\langle f(h)\rangle^2}\,,\label{eq:disc}
\end{equation}
which is an equal-time correlator between two different points in space. Note that because the mean field value vanishes, $\langle h\rangle = 0$, it would be incorrect to assume $\delta_h \propto \delta h/h$. 

As shown in Ref. \cite{Starobinsky:1994bd} (see also Refs. \cite{Motohashi:2012bb,Markkanen:2019kpv}), we can compute the equal-time correlator in terms of a spectral expansion of the unequal-time correlator 
\begin{equation}
\frac{\langle f(h(0))f(h(t))\rangle}{\langle f(h)\rangle^2}
=\sum_nf_n^2 e^{-\Lambda_n t}
\,,
\label{equ:spectral}
\end{equation}
where
\begin{equation}
f_n=\frac{\int {\rm d}h \psi_0(h)f(h)\psi_n(h)}{\int {\rm d} h \psi_0(h)f(h)\psi_0(h)}
\,,
\end{equation}
and $\Lambda_n$ and $\psi_n$ are the eigenvalues and orthonormal eigenvectors, respectively, of the eigenvalue equation
\begin{equation}
\bigg[\frac{1}{2}\frac{\partial^2}{\partial h^2}-\frac{1}{2}\left(v'(h)^2-v''(h)\right)\bigg]\psi_n(h)
=-\frac{4\pi^2\Lambda_n}{H^3}\psi_n(h)\,,\label{e:sch}
\end{equation} 
with 
\begin{equation}
    v(h)=\frac{4\pi^2}{3H^4}V(h) = \frac{\pi^2\lambda}{3}\left(\frac{h}{H}\right)^4.
\end{equation}

Because $f(h)$ is an even function, only even eigenvalues contribute to the spectral expansion. Also, because $\Lambda_0=0$, the
$n=0$ term cancels the disconnected part of the correlator \eqref{equ:spectral}. The leading non-trivial term is therefore to a good accuracy
\begin{equation}
\frac{\langle f(0)f(t)\rangle-\langle f\rangle^2 }{\langle f\rangle^2}
\approx
f_2^2 e^{-\Lambda_2t},
\label{equ:spectralapprox}
\end{equation}
and the higher order terms are exponentially suppressed. A numerical solution of the eigenvalue equation (\ref{e:sch}) gives
\begin{equation}
    \Lambda_2\approx 0.289\sqrt{\lambda}H.
\end{equation}
and
\begin{equation}
\label{eq:f2}
f_2(w)\approx 
\begin{cases}
0.796 & \quad w=0 ,\\
1.057 & \quad w=1/3 ,\\
1.268 & \quad w=1 .\\
\end{cases}
\end{equation}

Using de Sitter invariance we can then relate the unequal-time correlator (\ref{equ:spectral}) to the equal-time correlator with spatial separation by writing $t\rightarrow (2/H)\ln(aHr)$ \cite{Starobinsky:1994bd}. The correlator (\ref{eq:disc}) then becomes
\begin{equation}
\langle \delta_h(0)\delta_h(r) \rangle 
\approx
f_2^2(w)(aHr)^{-2\Lambda_2/H}\label{eq:num}\,,
\end{equation}
where $a$ and $H$ are to be evaluated at the end of inflation. The corresponding power spectrum can then be computed to be
\be
\label{eq:pp}
\mathcal{P}_{\delta h}= \mathcal{A}\bigg(\frac{k}{k_*}\bigg)^{n_h-1}\,,
\ee
where $k$ is the comoving wavenumber and $k_*=a_*H_*$ is a reference scale defined by horizon crossing at the time when the observable spectrum of perturbations was formed. In the following, we take $k_*=0.05\,{\rm Mpc}^{-1}$, as is customary. The amplitude and spectral index of the power spectrum \eqref{eq:pp} are
\begin{eqnarray}
\label{Ans}
\mathcal{A} &\simeq&  \frac{2f_2^2(w)}{\pi}\Gamma\left(2-(n_h - 1)\right)\sin\left(\frac{\pi(n_h - 1)}{2}\right)e^{-(n_h - 1)N_*} \nonumber\\
n_h &=&   1+2\Lambda_2 \simeq 1.579\sqrt{\lambda} ,
\end{eqnarray}
respectively. The number of e-folds $N_*$ between the end of inflation and horizon exit of the pivot scale $k_*$ depends on the scale of inflation, reheating temperature and the expansion history of the Universe after inflation and is given by (see e.g. Ref. \cite{Tenkanen:2019jiq})
\begin{eqnarray}
\label{efolds}
N_* &\simeq& 56  + \frac12\ln\left(\frac{r}{0.1}\right) -\ln\left(\frac{T_{\rm reh}}{10^{16}\,{\rm GeV}}\right) \\ \nonumber 
&+&\frac{1}{3(1+w)}\left(1.12+ 4\ln\left(\frac{T_{\rm reh}}{10^{16}\,{\rm GeV}}\right)
- \ln\left(\frac{r}{0.1}\right) \right) .
\end{eqnarray}

The amount of isocurvature perturbations can be described by defining the usual isocurvature parameter $\beta$ as
\be
\mathcal{P}_{S} =\frac{\beta}{1-\beta}\mathcal{P}_{\zeta},
\ee
which for correlated DM isocurvature is constrained by the Planck satellite to $\beta \lesssim 0.047$ at $k_*=0.05\,{\rm Mpc}^{-1}$, whereas $\mathcal{P}_{\zeta}(k_*)\simeq 2.1\times 10^{-9}$ \cite{Akrami:2018odb}. We can then write Eq.~(\ref{equ:PS}) as
\be
\label{isocurvaturelimit}
\beta \simeq \left(\left. \frac{3\rho_{\gamma}^{h}}{4\rho_{\gamma}^{H}}\right)^2\right|_{\rm reh} 
\frac{\mathcal{P}_{\delta h}}{\mathcal{P}_\zeta} ,
\ee
which gives the prediction of a given scenario for the isocurvature parameter $\beta$. Here the ratio between the energy density sourced by the Higgs field and the total energy density in the SM sector is evaluated at the time of reheating and we approximated $\rho_{\gamma}^{H} + \rho_{\gamma}^{h} \simeq \rho_{\gamma}^{H}$ but will consider below also the case where $\rho_{\gamma}^{h}$ gives a substantial contribution to the energy density already before reheating.

The ratio of the energy densities in Eq. \eqref{isocurvaturelimit} is given by
\be
\label{isocurvaturecontribution}
\left. \frac{\rho_{\gamma}^{h}}{\rho_{\gamma}^{H}}\right|_{\rm reh} 
= \frac{\lambda h_*^4}{12 H^2(a_{\rm reh})M_{\rm P}^2}\left(\frac{a_{\rm osc}}{a_{\rm reh}}\right)^4 ,
\ee
where $a_{\rm osc}$ denotes the time when the Higgs condensate began to oscillate, $3\lambda h_*^2=H^2(a_{\rm osc})$, the power $4$ for the ratio $a_{\rm osc}/a_{\rm reh}$ is due to the Higgs field and its decay products scaling as radiation, $H(a_{\rm reh}$ is given by Eq. \eqref{HandTreh}, and the scale factor at the time of reheating is $a_{\rm reh}=(H_*/H(a_{\rm reh}))^{2/(3(1+w))}$. For $h_*$ we use the typical value given by Eq. \eqref{hstar}. Trading then the scale of inflation $H_*$ with the tensor-to-scalar ratio $r$ as in Eq. \eqref{Hr} and substituting Eq. \eqref{isocurvaturecontribution} into Eq. \eqref{isocurvaturelimit}, we finally obtain a result for the isocurvature perturbations
\be
\label{rbetarelation}
\beta = C(w,\lambda, \mathcal{P}_\zeta, T_{\rm reh},r)r^{4-\frac{8}{3(1+w)}} ,
\ee
where $C$ is a function of the underlying parameters
\bea
C \equiv 2^{-12+\frac{8}{3(1+w)}}3^{-\frac{8}{3(1+w)}}\pi^{8-\frac{16}{3(1+w)}}A^{8-\frac{16}{3(1+w)}}B^{-4+\frac{16}{3(1+w)}} \nonumber\\
\times \lambda^{-\frac{4}{3(1+w)}}\mathcal{P}_\zeta^{3-\frac{32}{3(1+w)}}\left(\frac{T_{\rm reh}}{M_{\rm P}}\right)^{-8+\frac{32}{3(1+w)}}\mathcal{P}_{\delta_h}(r,T_{\rm reh},w) ,
\eea
where the factors $A$ and $B$ are the prefactors in Eqs. \eqref{hstar} and \eqref{HandTreh}, respectively, and $\mathcal{P}_{\delta_h}$ is given by Eq. \eqref{eq:pp}. 

The prediction for the amount of residual DM isocurvature perturbations, Eq. \eqref{rbetarelation}, is our main result, as it connects two observables, the isocurvature parameter $\beta$ and the primordial tensor-to-scalar ratio $r$ to each other in a simple and fairly model-independent way. In the next section, we will present the results for different cosmological parameters $w$ and $T_{\rm reh}$.

%

\section{Results}
\label{results}

\subsection{Radiation-dominated Universe}

In the standard radiation-dominated case where $w=1/3$, we obtain a simple result
\be
\label{rbeta}
\beta \simeq 10^{-13}\frac{r^2}{\sqrt{\lambda}} e^{-\sqrt{\lambda}\left(32.9+0.1\ln r \right)} ,
\ee
which for $\lambda \lesssim 10^{-3}$ simplifies to $\beta \simeq 10^{-13} r^2/\sqrt{\lambda}$. Therefore, given the observational upper limit on the tensor-to-scalar ratio $r\leq 0.06$ \cite{Ade:2018gkx}, the result \eqref{rbeta} shows that a detection of $\beta \gtrsim 10^{-15}$, i.e. in practice any value of $\beta$ observable in the foreseeable future, rules out all scenarios where $a)$ DM is a thermal relic, $b)$ there are no other sources of isocurvature perturbations besides the SM Higgs, and $c)$ the Universe was radiation-dominated after inflation. In the future the EUCLID satellite will constrain $\beta$ roughly at the percent level \cite{Amendola:2016saw}, whereas a detection of $r$ is possible at the level of $\mathcal{O}(10^{-2})$ by BICEP3 \cite{Wu:2016hul} or at $\mathcal{O}(10^{-3})$ by LiteBIRD \cite{Matsumura:2013aja}, CORE \cite{Remazeilles:2017szm} or the Simons Observatory \cite{Simons_Observatory}.

Therefore, the result shows that the SM Higgs cannot source observable DM isocurvature consistent with limits on $r$, and generating primordial DM isocurvature would require physics beyond the SM which includes a component that was not in thermal equilibrium with the DM nor the SM in the early Universe. Of course, interpreted in another way, we can conclude that the Higgs field is generically not a threat to models where both inflaton and DM reside in a decoupled hidden sector.

Note that the result is not sensitive to what drives inflation, when and how the inflaton field decays, nor what DM actually is, and in this sense the result is fairly model-independent. The result \eqref{rbeta} is inversely proportional to $\sqrt{\lambda}$ but given the very small number in front of $r^2/\sqrt{\lambda}$, the exact value of $\lambda$ does not matter unless the SM $\beta$-functions are severely fine-tuned and $\lambda \ll 1$ at the scale of inflation. Likewise, the result is not very sensitive to the average initial displacement of the Higgs field within our observable Universe today. As can be seen in Eq. \eqref{hstar}, the parameter $A$ characterizes how typical the field value and, consequently, the initial energy density in the SM sector were. The value of $\beta$ will not change in any appreciable way had the initial displacement been larger than that characterized by $A$: a deviation of $p$ standard deviations from the typical field value $h_* = \sqrt{\langle h^2\rangle}$ only changes the result by $p^{4-8/(3(1+w))}$, which is not enough to change our conclusion in any appreciable way. For example, for a $10\,\sigma$ deviation from the typical value used above, the result \eqref{rbeta} only changes by a factor $100$. Again, given the very small prefactor in Eq. \eqref{rbeta}, this will not change our conclusion.

\subsection{Matter-dominated Universe}

In a matter-dominated Universe with $w=0$, we obtain
\begin{eqnarray}
\label{rbetaMD}
\beta &\simeq& 10^{-13}\left(\frac{T_{\rm reh}}{10^{16}\,{\rm GeV}}\right)^{8/3}\frac{r^{4/3}}{\lambda^{5/6}} \nonumber\\
&\times& e^{-\sqrt{\lambda}\left(32.9+0.1\ln r + 0.2\ln\left(\frac{T_{\rm reh}}{10^{16}\,{\rm GeV}}\right)\right)} .
\end{eqnarray}
We see again that a discovery of a non-zero $\beta$ in any foreseeable future rules out this type of scenarios because they predict negligible $\beta$. Again, at the same time the result shows that residual DM isocurvature, despite being unavoidably generated, is not a problem for this type of scenarios, as studied recently in Refs. \cite{Berlin:2016vnh,Tenkanen:2016jic,Berlin:2016gtr,Heurtier:2019eou}. Note that the result \eqref{rbetaMD} does not exactly coincide with Eq. \eqref{rbeta} at the limit of maximum reheating temperature for this scenario $T_{\rm reh}/M_{\rm P}\simeq 0.004r^{1/4}\lambda^{1/8}$, as given in Eq. \eqref{Trehupper2}. This is because the solutions for the eigenfunctions \eqref{eq:f2} and number of e-folds \eqref{efolds} assume a discrete value for $w$, and changing $T_{\rm reh}$ only is not sufficient to go to the limit of prompt reheating. However, for $\lambda \ll 1$ the exponential in Eq. \eqref{rbetaMD} goes to unity and the MD and RD results coincide to a good accuracy at the limit $T_{\rm reh}/M_{\rm P}\to 0.004r^{1/4}\lambda^{1/8}$.

\subsection{Kination-dominated Universe}
\label{sec:KD}

The above results are quite natural in a sense that when the Higgs condensate and its decay products scale as $a^{-4}$ and $w\leq 1/3$, they cannot increase their fraction of the total energy density. Let us therefore now discuss a case where $w>1/3$ and the SM sector started to increase its fractional energy density already before the hidden sector decayed into the SM. We consider two cases: one in which the SM sector remained strictly subdominant until reheating so that Eq. \eqref{isocurvaturecontribution} still holds, and one where it became the dominant energy density component early on. This is possible if the Universe underwent e.g. a {\it kination} phase where the total energy density after inflation was dominated by a fast-rolling inflaton field with $w=1$, making the hidden sector energy density dilute as $a^{-6}$ \cite{Spokoiny:1993kt,Joyce:1996cp}.

\begin{figure}
\begin{center}
\includegraphics[width=.495\textwidth]{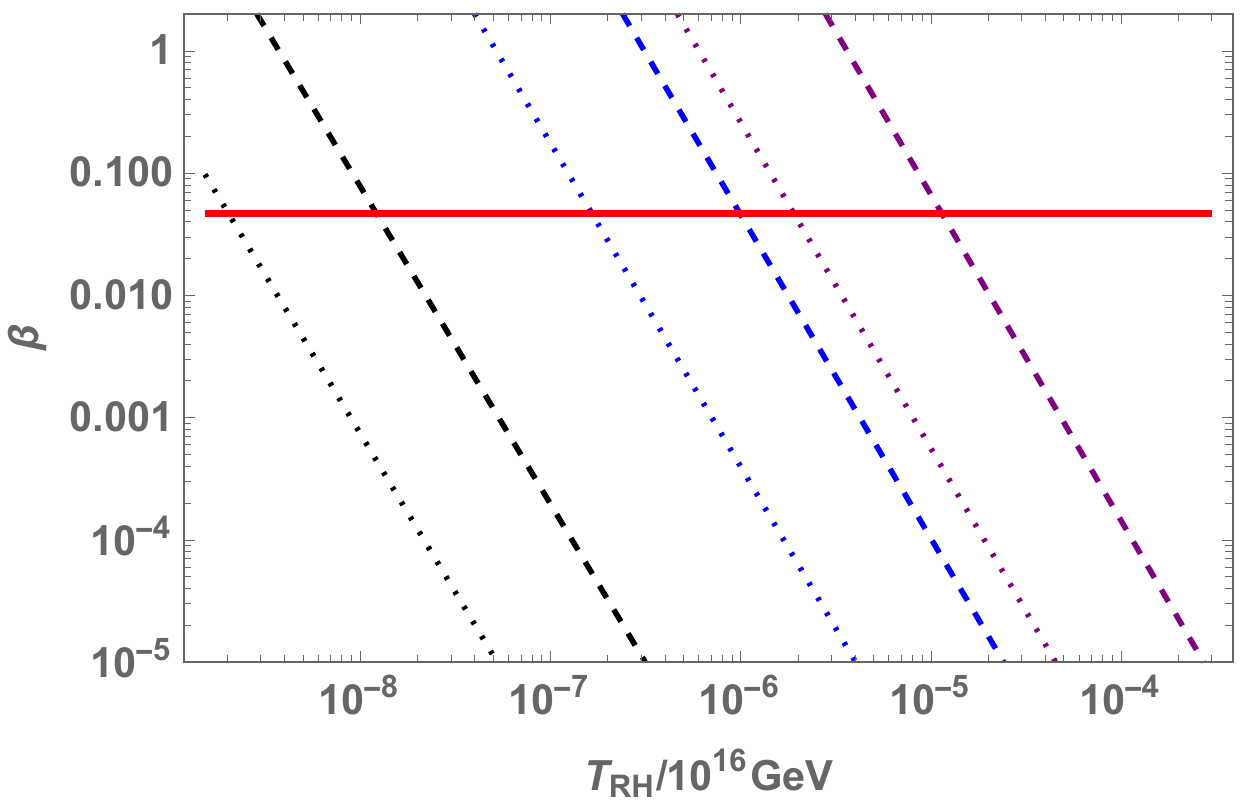}
\caption{The isocurvature parameter $\beta$ as a function of reheating temperature $T_{\rm reh}$ in the $w=1$ case. For black/blue/purple lines (from left to right) $\lambda=0.13, 0.01, 10^{-8}$ and $r=0.06,0.01$ for dashed/dotted lines, respectively. The horizontal red line is the observational upper limit $\beta=0.047$ \cite{Akrami:2018odb}. The isocurvature parameter is negligible in the $w=0,1/3$ cases.}
\label{beta_vs_Trh}
\end{center}
\end{figure}

Thus, in the case of $w=1$, we obtain
\begin{eqnarray}
\label{rbetaKD}
\beta &\simeq& 10^{-13}\left(\frac{T_{\rm reh}}{10^{16}\,{\rm GeV}}\right)^{-8/3}\frac{r^{8/3}}{\lambda^{1/6}} \nonumber\\
&\times& e^{-\sqrt{\lambda}\left(33.0+0.2\ln r - 0.2\ln\left(\frac{T_{\rm reh}}{10^{16}\,{\rm GeV}}\right)\right)} ,
\end{eqnarray}
when the Higgs and its decay products remain energetically subdominant up until reheating. Notably, in Eq. \eqref{rbetaKD} the terms proportional to $T_{\rm reh}$ come with a sign different from those in Eq. \eqref{rbetaMD}, reflecting the fact that in this scenario the SM sector is increasing its fractional energy density and, consequently, increasing the residual DM isocurvature perturbation. Again, as with the MD case, the result \eqref{rbetaKD} does not exactly coincide with the result \eqref{rbeta} at the limit of maximum reheating temperature unless $\lambda \ll 1$. Also, the result \eqref{rbetaKD} only holds for a sufficiently large reheating temperature given by
\be
\label{Trehlower}
\frac{T_{\rm reh}}{10^{16}\,{\rm GeV}} > 7.3\,\mathcal{P}_\zeta \lambda^{-1/4}r ,
\ee
which results from requiring that the left-hand side of Eq. \eqref{isocurvaturecontribution} remains smaller than unity. The result \eqref{rbetaKD} is shown in Fig. \ref{beta_vs_Trh} for various choices of parameters. The lower limit for $T_{\rm reh}$ as given by Eq. \eqref{Trehlower} varies between $10^7$ GeV and $10^9$ GeV for the contours shown but is always below the value for which $\beta$ becomes substantial, as well as the maximum reheating temperature \eqref{Trehupper2}. The results show that if the hidden sector underwent a kination phase, the presence of the Higgs field can generate significant residual DM isocurvature despite the fact that the SM sector remains a subdominant component until reheating.

At the opposite limit where the SM sector becomes to dominate the total energy density before the hidden sector decays, the SM radiation is reheated by the Higgs only. Hence, the SM Higgs will in this case act as a curvaton \cite{Enqvist:2001zp,Lyth:2001nq,Moroi:2001ct}, imprinting its fluctuation spectrum into metric perturbations. This is the scenario studied in Refs. \cite{Kunimitsu:2012xx,Figueroa:2016dsc} (see also Ref. \cite{DeSimone:2012qr}), where the conclusion was that this scenario must be ruled out because the Higgs has a roughly scale-invariant fluctuation spectrum with a very large amplitude. However, as Eq. \eqref{Ans} clearly shows, fluctuations in the Higgs' energy density are not scale-invariant (unless $\lambda$ is tiny), and hence the spectrum can, in fact, have a small enough amplitude at large scales to coincide with the observed value $\mathcal{P}_\zeta(k_*) \simeq 2.1\times 10^{-9}$. However, the success of the scenario is unavoidably spoiled by the {\it blue} tilt of the Higgs' perturbation spectrum, as observations of the CMB show that the spectrum is {\it red-tilted} at the $8\,\sigma$ confidence level \cite{Aghanim:2018eyx}. Therefore, this scenario is ruled out.

Before concluding this section, let us make a remark on the magnitude of perturbations at smaller scales. Because the isocurvature spectrum is very blue-tilted, the perturbations can become large at small scales and even exceed the initial curvature perturbations if their spectrum has a small (red) tilt. Even though the Higgs cannot act as a traditional curvaton field sourcing the perturbations at large scales, it can still do so at small scales. By defining a curvature perturbation on the uniform energy density hypersurface of each fluid $i$ as $\zeta_i \equiv -\Phi -H\delta\rho_i/\dot{\rho_i}$ \cite{Wands:2000dp}, it can be shown that in this case the total curvature perturbation is at early times
\begin{equation}
\zeta = \frac{\rho_\gamma^h\zeta_\gamma + \frac32\rho_{\rm hid}\zeta_{\rm hid}}{\rho_h^\gamma + \frac32\rho_{\rm hid}} ,
\end{equation}
which shows that the Higgs can give an appreciable contribution to the curvature perturbation at scales where $\zeta_h/\zeta_{\rm hid} \sim \rho_{\rm hid}/\rho_\gamma^h$, where $\rho_{\rm hid}/\rho_\gamma^h$ is to be evaluated at the decay of the hidden sector. Because $\zeta_\gamma \propto -H\delta\rho_i/\dot{\rho_i} \propto \delta\rho_\gamma^h/\rho_\gamma^h$, the curvature perturbation sourced by the Higgs can be computed using Eq. \eqref{perturbations_higgs}. However, as $\zeta_{\rm hid}$ depends on the inflationary sector, a more detailed investigation of this possibility is beyond the scope of this paper.


\section{Conclusions}
\label{conclusions}

In this paper we studied scenarios where the inflaton field decays dominantly to a hidden sector which is thermally decoupled from the visible SM sector. By assuming that the DM fluid inherits the same perturbation spectrum as the inflaton and the comoving DM number density freezes before the eventual decay of the hidden sector and utilizing the typical behavior of the SM Higgs during inflation, we derived a relation between the primordial tensor-to-scalar ratio $r$ and amplitude of the DM isocurvature perturbations $\beta$, Eq. \eqref{rbetarelation}. 

We considered different expansions histories and found that in the standard case where the Universe was radiation-dominated after inflation, the observables are connected by $\beta\sim 10^{-13}r^2/\sqrt{\lambda}$, whereas in matter- and kination-dominated cases there is extra dependence on the reheating temperature, as depicted by Eqs. \eqref{rbetaMD} and \eqref{rbetaKD}. The above result for the radiation-dominated case applies for small Higgs self-coupling, $\lambda \ll 1$, and the full result can be found in Eq. \eqref{rbeta}. These equations constitute our main results together with the conclusion that the Higgs field cannot source the curvature perturbation at large scales, i.e. act as a curvaton field, because of its blue-tilted fluctuation power spectrum. However, the Higgs can still give a sizeable contribution to the curvature perturbation at smaller scales, as discussed in Sec. \ref{sec:KD}.

We conclude that a future discovery of primordial DM isocurvature of practically any magnitude will rule out all simple scenarios where DM is a thermal relic either from the visible sector or from a hidden one and where the Universe was either radiation- or matter-dominated before reheating of the SM. At the same time, the results show that the Higgs field is generically not a threat to models where both the inflaton and DM reside in a decoupled hidden sector. However, a phase of an early kination-domination provides for an exception to these conclusions, as then large DM isocurvature can be generated from the SM Higgs even in cases where the SM sector remained subdominant until reheating. The results underline the importance of studying not only the evolution of fractional energy densities of different sectors but also how the perturbations in them evolved in the early Universe.


\section*{Acknowledgements}
I thank Marc Kamionkowski, Karim Malik, Tommi Markkanen, David Mulryne, Sami Nurmi, and Arttu Rajantie for useful discussions. I also acknowledge the Simons Foundation for funding and Imperial College London and Helsinki Institute of Physics for hospitality.

\bibliography{HH_isocurvature.bib}

\end{document}